\documentclass[prd,showpacs,preprintnumbers,amsmath,amssymb]{revtex4}
\usepackage{graphicx}% Include figure files
\usepackage{dcolumn}% Align table columns on decimal point
\usepackage{bm}% bold math
\def\be{\begin{eqnarray}}
\def\ee{\end{eqnarray}}
\def\bea{\begin{eqnarray}}
\def\eea{\end{eqnarray}}

\def\0T{{\bf 0}_\perp}
\begin{document}

%\preprint{draft}

\title{Angular Momentum Decomposition for an Electron}
% Force line breaks with \\

\author{Matthias Burkardt\footnote{present address:
Thomas Jefferson National Accelerator Facility, Newport News, 
VA 23606, U.S.A.} and Hikmat BC}%, Abdullah Jarrah}
 \affiliation{Department of Physics, New Mexico State University,
Las Cruces, NM 88003-0001, U.S.A.}
%\date{\today}% It is always \today, today,
%             %  but any date may be explicitly specified

\begin{abstract}
We calculate the orbital angular momentum
of the `quark' in the scalar diquark model as well as that of the
electron in QED (to order $\alpha$). We compare the orbital angular 
momentum obtained
from the Jaffe-Manohar decomposition to that obtained from
the Ji relation and estimate the importance of the vector potential 
in the definition of orbital angular momentum. 
%In the framework of 
%the bag model, we also estimate the effect that the spectators may
%have in this context.
\end{abstract}

%\pacs{Valid PACS appear here}% PACS, the Physics and Astronomy
                             % Classification Scheme.
%\keywords{Suggested keywords}%Use showkeys class option if keyword
                              %display desired
\maketitle
\narrowtext
\section{Introduction}
While the total angular momentum  of an isolated system is 
uniquely defined, ambiguities arise when decomposing the total angular
momentum of an interacting multi-constituent system into
contributions from various constituents. Moreover, in a gauge 
theory, switching the gauge may result in shuffling angular momentum
between matter and gauge degrees of freedom.
In the context of nucleon structure, this gives rise to subtleties
in defining these quantities that are more fundamental than those
subtleties associated with the choice of factorization scheme.

In the context of hadron structure, it is natural to perform a 
decomposition of the $\hat{z}$ component of the angular momentum as 
the $\hat{z}$ component of the quark spin has a partonic 
interpretation as a difference between parton densities. Indeed, in 
the light-cone framework, Jaffe and Manohar proposed a decomposition 
of the form \cite{JM}
\be
\frac{1}{2}=\frac{1}{2}\sum_q\Delta q + \sum_q {\cal L}_q^z+
\frac{1}{2}\Delta G + {\cal L}_g^z,
\label{eq:JJM}
\ee
whose terms are defined as matrix elements of the corresponding
terms in the $+12$ component of the angular momentum tensor
\be
M^{+12} = \frac{1}{2}\sum_q q^\dagger_+ \gamma_5 q_+ +
\sum_q q^\dagger_+\left({\vec r}\times i{\vec \partial}
\right)^z q_+  
+ \varepsilon^{+-ij}\mbox{Tr}F^{+i}A^j
+ 2 \mbox{Tr} F^{+j}\left({\vec r}\times i{\vec \partial} 
\right)^z A^j.
\label{M+12}
\ee
The first and third term in (\ref{eq:JJM},\ref{M+12}) are the
`intrinsic' contributions 
(no factor of ${\vec r}\times $) to the nucleon's
angular momentum $J^z=+\frac{1}{2}$ and have a 
physical interpretation as quark and gluon spin respectively, while
the second and fourth term can be identified with the quark/gluon
orbital angular momentum (OAM).
Here $q_+ \equiv \frac{1}{2} \gamma^-\gamma^+ q$ is the dynamical
component of the quark field operators, and light-cone gauge
$A^+\equiv A^0+A^z=0$ is implied. 
The residual gauge invariance is fixed by
imposing anti-periodic boundary conditions 
${\bf A}_\perp({\bf x}_\perp,\infty^-)=-
{\bf A}_\perp({\bf x}_\perp,-\infty^-)$ on the transverse components
of the vector potential.

Since the quark spin term does not contain any derivatives,
its manifest gauge invariance is evident. However, $\Delta G$ is also
gauge invariant, as it is experimentally accessible. In gauges
other than light-cone gauge, it is defined through a non-local 
operator \cite{BJ}. The net parton OAM
\be
{\cal L}^z=\sum_q {\cal L}_q^z +{\cal L}_g^z = \frac{1}{2} - 
\frac{1}{2}\sum_q\Delta q - \frac{1}{2}\Delta G
\ee
can be related to differences between observables and is thus
also obviously gauge invariant. However, similar to the case of
$\Delta G$, a manifestly gauge invariant operator defining 
${\cal L}^z$
would be non-local, reducing to a local expression in light-cone
gauge only.
For the individual OAMs
the situation is more subtle and a detailed discussion can be 
found in Ref. \cite{BJ}.

An alternative decomposition \cite{JiPRL} of the nucleon spin 
\be
\frac{1}{2}=\frac{1}{2}\sum_q\Delta q + \sum_q { L}_q^z+
J_g^z
\label{eq:JJi}
\ee
into quark spin, quark OAM, and gluon (total) angular momentum
is obtained from the expectation value of 
\be
M^{0xy}= \sum_q \frac{1}{2}q^\dagger \Sigma^zq +
\sum_q q^\dagger \left({\vec r} \times i{\vec D}
\right)^zq
+  
\left[{\vec r} \times \left({\vec E} \times {\vec B}\right)\right]^z
\label{M012}
\ee
with $i{\vec D}=i{\vec \partial}-g{\vec A}$.
Its main advantages are that each term can be expressed as the
expectation value of a manifestly gauge invariant
local operator and that the
quark total angular momentum $J_q^z=\frac{1}{2}\Delta q+L_q^z$
can be related to generalized parton distributions (GPDs), using
\cite{JiPRL} 
\be
J_q^z = \frac{1}{2}\int_0^1 dx\,x\left[q(x)+E_q(x,0,0)\right],
\label{eq:Jirelation} 
\ee
and can thus be measured in deeply virtual Compton scattering or 
calculated in lattice gauge theory.
Its main disadvantage is that both quark OAM 
$L_q^z$ as well as gluon angular momentum $J_g^z$ contain
interactions through the vector potential in the gauge covariant
derivative, which complicates their physical interpretation.

Since the expectation value of $\bar{q}\gamma^z \Sigma^zq$ vanishes
for a parity eigenstate, one can replace 
$q^\dagger \Sigma^zq \longrightarrow \bar{q} \gamma^+ \Sigma^zq
=q_+^\dagger \gamma_5 q_+$, i.e. the $\Delta q$ are common
to both decompositions. This is not the case for all the other
terms. For example, the angular momenta in these decompositions 
(\ref{eq:JJM}),(\ref{eq:JJi}) are not defined through matrix elements
of the same operator and one should not expect them
to have the same numerical value. However, no intuition exists as to 
how large that difference is.
\begin{figure}
\unitlength1.cm
\begin{picture}(10,6)(3.5,15)
\includegraphics{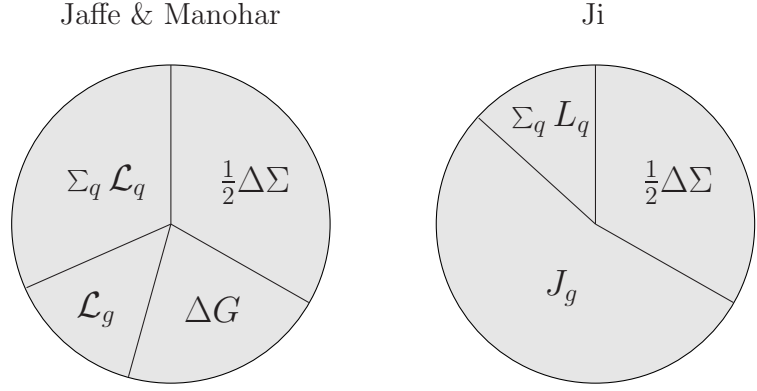}
\end{picture}
\caption{Schematic comparison between the two decompositions
(\ref{eq:JJM}) and (\ref{eq:JJi}) of the nucleon spin.
In general, only 
$\frac{1}{2}\Delta \Sigma \equiv \frac{1}{2}\sum_q\Delta q$ 
is common to both decompositions.}
\end{figure}

In the matrix element defining $L_q^z$, one may make the replacement
\be
q^\dagger \left({\vec r}\times i{\vec D}\right)^zq
= \bar{q}\gamma^0\left({\vec r}\times i{\vec D}\right)^zq
\longrightarrow \bar{q}\left(\gamma^0+\gamma^z\right)
\left({\vec r}\times i{\vec D}\right)^zq
= q_+^\dagger \left({\vec r}\times i{\vec D}\right)^zq_+,
\label{eq:replace}
\ee
provided that the expectation value is taken in a parity eigenstate.
While the Dirac structure of the operator on the r.h.s. of 
(\ref{eq:replace}) is now the same as that appearing in (\ref{M+12}),
Eq. (\ref{eq:replace}) still contains the transverse component of the
vector potential through the gauge covariant derivative, 
and therefore, even in light-cone gauge,
${\cal L}_q^z$ and $L_q^z$ differ by the expectation value of
$q_+^\dagger \left({\vec r}\times g{\vec A}\right)^zq_+$.
While it has long been realized that in general 
${\cal L}_q^z\neq L_q^z$,  The main purpose of this paper is to 
address
this issue first in the context of a scalar diquark model and then
in QED.

\section{Orbital Angular Momentum in the Scalar
Diquark Model}
In a two particle system we introduce center of momentum and
relative $\perp$ coordinates as
\be
{\bf P}_\perp &\equiv& {\bf p}_{1\perp}+{\bf p}_{2\perp}
\label{eq:CMrel}\\
{\bf R}_\perp &\equiv& x_1{\bf r}_{1\perp}+ x_2{\bf r}_{2\perp}
= x{\bf r}_{1\perp}+(1-x){\bf r}_{2\perp}\nonumber\\
{\bf k}_\perp &\equiv& x_2{\bf p}_{1\perp}-x_1{\bf p}_{2\perp}=
(1-x){\bf p}_{1\perp}-x{\bf p}_{2\perp}\nonumber\\
{\bf r}_\perp &\equiv& {\bf r}_{1\perp}
-{\bf r}_{2\perp}
\ee
where $x_1=x$ and $x_2=1-x$ are the momentum fractions carried by
the active quark and the spectator respectively.
For a state with ${\bf P}_\perp=0$, this implies ${\bf p}_{1\perp}
=-{\bf p}_{2\perp}={\bf k}_\perp$, allowing one to 
replace the OAM operator for particle 1 by ($1-x$) times the relative
OAM in such a state  \cite{hari}
\be
{\cal L}_1^z = {\bf r}_{1\perp} \times {\bf p}_{1\perp}
= \left[ {\bf R}_\perp + (1-x){\bf r}_\perp\right] \times
{\bf k}_\perp \longrightarrow (1-x){\bf r}_\perp \times {\bf k}_\perp
= (1-x) {\cal L}^z .
\label{eq:Lz1}
\ee
Here we used that the internal wave function of a bound state
satisfies $\langle {\bf k}_\perp \rangle=0$. Likewise one
finds that the expectation value of ${\cal L}_2^z$ can be
replaced by the expectation value of $x {\cal L}^z$.

We now use the above decompositions (\ref{eq:JJM}),(\ref{eq:JJi})
to calculate the OAM of the
`quark' in the scalar diquark model, where the
two particle Fock space amplitudes read \cite{ELz}
\be
\psi_{+\frac{1}{2}}^\uparrow \left(x,{\bf k}_\perp\right)
&=& \left(M+\frac{m}{x}\right) \phi (x,{\bf k}_\perp^2) 
\label{eq:SDQM}\\
\psi_{-\frac{1}{2}}^\uparrow (x,{\bf k}_\perp)
&=&-\frac{k^1+ik^2}{x} \phi (x,{\bf k}_\perp^2)
\nonumber\ee
with $\phi = \frac{g/\sqrt{1-x}}{M^2-\frac{{\bf k}_\perp^2+m^2}{x}
-\frac{{\bf k}_\perp^2+\lambda^2}{1-x}}$.
Here $g$ is the Yukawa coupling and $M$/$m$/$\lambda$ are the masses 
of the `nucleon'/`quark'/diquark respectively. Furthermore
$x$ is the momentum 
fraction carried by the quark and ${\bf k}_\perp\equiv 
{\bf k}_{\perp e}-
{\bf k}_{\perp \gamma}$ represents the relative $\perp$ momentum.
The upper wave function index
$\uparrow$ refers to the helicity of the `nucleon' and the
lower index to that of the quark. With the light-cone
wave functions available (\ref{eq:SDQM}), 
it is straightforward to compute
either ${\cal L}_q^z$ or $J_q^z$, and hence $L_q^z$ from the Ji
relation.

This yields for the orbital angular momentum ${\cal L}_q^z$ of the 
`quark' 
\be
{\cal L}_q^z = \int_0^1 dx \int \frac{d^2{\bf k}_\perp}{16\pi^3}
(1-x) \left|\psi_{-\frac{1}{2}}^\uparrow\right|^2.
\label{eq:calLSDQM}
\ee
Alternatively one may consider the OAM as obtained from GPDs using
the Ji relation (\ref{eq:Jirelation}) as
\be
L_q^z = \frac{1}{2}\!\int_0^1\!dx\, 
\left[xq(x)+xE(x,0,0)-\Delta q (x) \right],
\label{eq:LSDQM}
\ee
where
\be
xq(x)&=&Z\delta(1-x) + x  \int \frac{d^2{\bf k}_\perp}{16\pi^3}
\left[ \left|\psi_{+\frac{1}{2}}^\uparrow\right|^2 +
\left|\psi_{-\frac{1}{2}}^\uparrow\right|^2\right]
\label{eq:pdfSDQM}\\
\Delta q(x)&=&Z\delta(1-x) + \int \frac{d^2{\bf k}_\perp}{16\pi^3}
\left[ \left|\psi_{+\frac{1}{2}}^\uparrow\right|^2 -
\left|\psi_{-\frac{1}{2}}^\uparrow\right|^2\right]\nonumber\\
xE(x,0,0)&=& 2Mg^2x\int \frac{d^2{\bf k}_\perp}{16\pi^3}\frac{
(1-x)^2\left(xm+M\right)}
{\left[x(1-x)M^2 - (1-x)m^2 -x \lambda^2 - {\bf k}_\perp^2\right]^2}
= \frac{Mg^2}{8\pi^2}\frac{x(1-x)^2\left(xm+M\right)}{-x(1-x)M^2 + (1-x)m^2 +x\lambda^2}.
\nonumber
\ee
As one may have expected, the wave function renormalization constant
\be
Z=1-\int_0^1dx \int \frac{d^2{\bf k}_\perp}{16\pi^3}
\left[ \left|\psi_{+\frac{1}{2}}^\uparrow\right|^2 +
\left|\psi_{-\frac{1}{2}}^\uparrow\right|^2\right]
\ee
cancels in $L_q^z$, yielding
\be
\label{eq:LSDQM2}
L_q^z = \frac{1}{2}\int_0^1 dx\int \frac{d^2{\bf k}_\perp}{16\pi^3}
\left[
(x-1)\left|\psi_{+\frac{1}{2}}^\uparrow\right|^2
+(x+1)\left|\psi_{-\frac{1}{2}}^\uparrow\right|^2\right]
+\frac{1}{2}\int_0^1 dx xE(x,0,0).
\ee
Since some of the above ${\bf k}_\perp$ integrals diverge, 
a manifestly Lorentz invariant
Pauli-Villars regularization (subtraction with heavy scalar
$\lambda^2\rightarrow \Lambda^2$) is always understood.
Evaluating the above integrals is tedious, but straightforward, and
one finds
\be
{\cal L}_q^z = L_q^z
\ee
as was expected since $L_q^z$ in the scalar diquark model does
not contain a gauge field term.
\begin{figure}
\unitlength1cm
\begin{picture}(15,9)(4,17)
\includegraphics{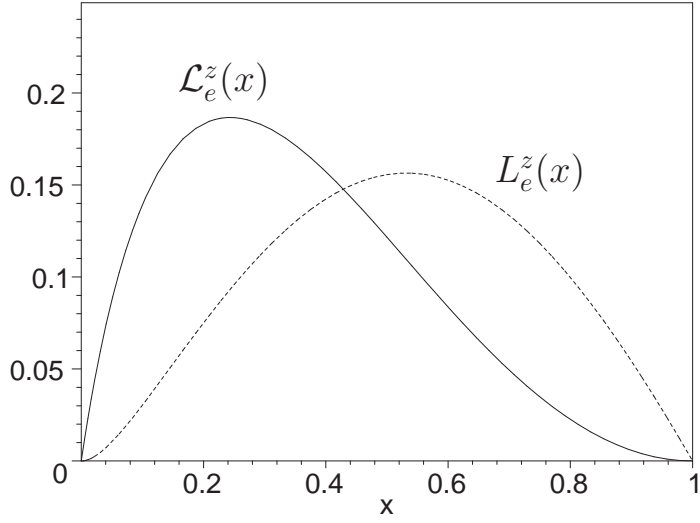}
\end{picture}
\caption{$x$ distribution of the orbital angular momentum
${\cal L}_q^z(x)$ (full) compared to $L_q^z(x)$ from the unintegrated
Ji relation (dotted) in the scalar diquark model
for parameters $\Lambda^2=10m^2=10\lambda^2$.
Both in units of $\frac{g^2}{16\pi^2}$.} 
%(the figure is on the last page!)}
\label{fig:L(x)}
\centering
\end{figure}
However, there is no such identity for the OAM distribution.
The distribution of the $\hat{z}$ component of the OAM 
${\cal L}_q^z(x)$ is defined as in (\ref{eq:calLSDQM}), but without
the $x$-integration. A comparison with 
(\ref{eq:LSDQM}) without $x$-integration, i.e. comparing 
${\cal L}_q^z(x)$ with $L_q^z(x)\equiv
\frac{1}{2}\left[xq(x)+xE(x,0,0)-\Delta q(x)\right]$ 
(Fig. \ref{fig:L(x)})
shows that, 
even in a model without gauge fields, $L_q^z(x)$
cannot 
be identified with the $x$-distribution of ${\cal L}_q^z$ 
for a longitudinally polarized nucleon \cite{hoodbhoy}.

\section{Orbital Angular Momentum in QED}
In QED, there are four polarization states in the $e\gamma$
Fock component. To lowest order, the respective Fock space
amplitudes for a dressed electron with $J^z=+\frac{1}{2}$ read
\be
\Psi^\uparrow_{+\frac{1}{2}+1}(x,{\bf k}_\perp)
&=& \frac{k^1-ik^2}{x(1-x)}\phi(x,{\bf k}_\perp^2)
\\
\Psi^\uparrow_{+\frac{1}{2}-1}(x,{\bf k}_\perp)
&=& -\frac{k^1+ik^2}{1-x}\phi(x,{\bf k}_\perp^2)\nonumber\\
\Psi^\uparrow_{-\frac{1}{2}+1}(x,{\bf k}_\perp)
&=& \left(\frac{m}{x}-m\right)\phi(x,{\bf k}_\perp^2) \nonumber\\
\Psi^\uparrow_{-\frac{1}{2}-1}(x,{\bf k}_\perp)&=&0\nonumber
\ee
with $\phi(x,{\bf k}_\perp^2) = \frac{\sqrt{2}}{\sqrt{1-x}}\frac{e}{M^2-\frac{{\bf k}_\perp^2+m^2}{x}
-\frac{{\bf k}_\perp^2+\lambda^2}{1-x}}$.

Using these light-cone wave functions, it is again straightforward to
calculate the orbital angular momentum (\ref{eq:Lz1}) 
of the electron in the Jaffe-Manohar \cite{JM} decomposition
\be
{\cal L}_e^z= \int_0^1dx\int \frac{d^2{\bf k}_\perp}{16\pi^3}
(1-x)\left[
\left|\Psi^\uparrow_{+\frac{1}{2}-1}(x,{\bf k}_\perp)
\right|^2 -\left|\Psi^\uparrow_{+\frac{1}{2}+1}(x,{\bf k}_\perp)
\right|^2 \right]
\label{eq:JMQED}
\ee
Likewise, it is straightforward to evaluate the OAM using the Ji
relation
\be
L_e^z = \frac{1}{2}\int_0^1dx\, \left[x q_e(x)+xE_e(x,0,0)-
\Delta q_e(x)\right]
\label{eq:JiQED}
\ee
with \cite{ELz}
\be
xq_e(x)&=&Z\delta(1-x) + x  \int \frac{d^2{\bf k}_\perp}{16\pi^3}
\left[ \left|\psi_{+\frac{1}{2},+1}^\uparrow\right|^2 +
\left|\psi_{+\frac{1}{2},-1}^\uparrow\right|^2 +
\left|\psi_{-\frac{1}{2},+1}^\uparrow\right|^2\right]
\label{eq:pdfVDQM}\\
\Delta q_e(x)&=&Z\delta(1-x) + \int \frac{d^2{\bf k}_\perp}{16\pi^3}
\left[ \left|\psi_{+\frac{1}{2},+1}^\uparrow\right|^2+
\left|\psi_{+\frac{1}{2},-1}^\uparrow\right|^2 -
\left|\psi_{-\frac{1}{2},+1}^\uparrow\right|^2\right]\nonumber\\
xE_e(x,0,0) &=& 4m^2e^2 \int \frac{d^2{\bf k}_\perp}{16\pi^3}
\frac{x^2(1-x)^2}{\left[m^2(1-x)^2+\lambda^2 x+{\bf k}_\perp^2
\right]^2} = \frac{m^2e^2}{4\pi^2}\frac{x^2(1-x)^2}{m^2(1-x)^2+
\lambda^2 x}.\nonumber
\ee
Again the wave function renormalization constant
\be
Z=1-\int_0^1 dx\int \frac{d^2{\bf k}_\perp}{16\pi^3}
\left[ \left|\psi_{+\frac{1}{2},+1}^\uparrow\right|^2+
\left|\psi_{+\frac{1}{2},-1}^\uparrow\right|^2 +
\left|\psi_{-\frac{1}{2},+1}^\uparrow\right|^2\right]
\ee
drops out in (\ref{eq:JiQED}), yielding
\be
L_e^z = \frac{1}{2}\int_0^1 dx\int \frac{d^2{\bf k}_\perp}{16\pi^3}
\left[(x-1)\left|\psi_{+\frac{1}{2},+1}^\uparrow\right|^2+
(x-1)\left|\psi_{+\frac{1}{2},-1}^\uparrow\right|^2+
(x+1)\left|\psi_{-\frac{1}{2},+1}^\uparrow\right|^2\right]
+ \frac{1}{2}\int_0^1 dx xE_e(x,0,0).
\ee
Because of the divergent ${\bf k}_\perp$ integrals a Pauli-Villars
subtraction with $\lambda^2\longrightarrow \Lambda^2$ is understood 
and
$\lambda^2\longrightarrow 0$ at the end of the calculation, while
$\Lambda^2 \gg m^2$.

\begin{figure}
\unitlength1cm
\begin{picture}(15,9.0)(4,17)
\includegraphics{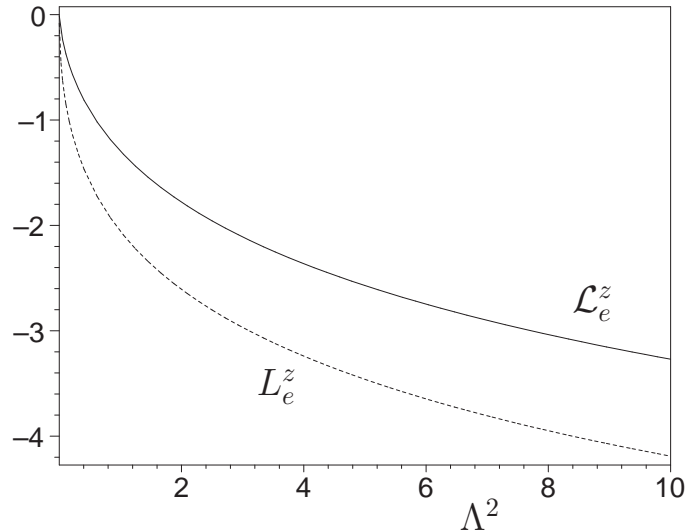}
\end{picture}
\caption{Cutoff dependence of ${\cal L}_e^z$ (full) and
$L_e^z$ (dotted). Both in units of $\frac{\alpha}{4\pi}$.}
\label{fig:cutoff}
\centering
\end{figure}

The evaluation of the above integrals is again straightforward,
yielding
\be
{\cal L}_e^z = -\frac{\alpha}{2\pi} \int_0^1 dx (1-x^2)\log
\frac{(1-x)^2m^2+x\Lambda^2}{(1-x)^2m^2+x\lambda^2}
\stackrel{\stackrel{\Lambda \rightarrow \infty}{\lambda \rightarrow 0}}{\longrightarrow}
-\frac{\alpha}{4\pi}\left[\frac{4}{3}\log \frac{\Lambda^2}{m^2}
-\frac{2}{9}\right]
\ee
and
\be
L_e^z & &=-\frac{\alpha}{4\pi}\int_0^1 dx (1+x^2)\left[\log
\frac{(1-x)^2m^2+x\Lambda^2}{(1-x)^2m^2+x\lambda^2}
-\frac{(1-x)^2m^2}{(1-x)^2m^2+x\lambda^2}
+\frac{(1-x)^2m^2}{(1-x)^2m^2+x\Lambda^2}\right]\\
& &\stackrel{\stackrel{\Lambda \rightarrow \infty}{\lambda \rightarrow 0}}{\longrightarrow}-\frac{\alpha}{4\pi}\left[
\frac{4}{3}\log \frac{\Lambda^2}{m^2}+\frac{7}{9}\right].\nonumber
\ee
Both ${\cal L}_e^z$ and $L_e^z$ are negative, regardless of the
value of $\Lambda^2$ (as long as $\Lambda^2>\lambda^2$).
In the case of ${\cal L}_e^z$ the physical reason is helicity
retention \cite{BBS}, 
which favors the emission of photons with the spin parallel 
(as compared to anti-parallel)
to the original quark spin --- particularly when $x\rightarrow 0$
--- resulting more likely in a state with negative OAM.
The divergent parts of ${\cal L}_e^z$ and $L_e^z$ are the same so 
that their difference is UV finite (Fig. \ref{fig:cutoff})
\be
{\cal L}_e^z - L_e^z \stackrel{\stackrel{\Lambda \rightarrow \infty}{\lambda \rightarrow 0}}{\longrightarrow} 
\frac{\alpha}{4\pi}. \label{eq:LLQED}
\ee

Applying these results to a (massive) quark with
$J^z=+ \frac{1}{2}$ yields to ${\cal O}(\alpha_s)$
\be
{\cal L}_q^z - L_q^z = \frac{\alpha_s}{3\pi},
\ee
i.e., for $\alpha_s \approx 0.5$ about 10\% of the spin budget for
this quark.

In QCD, the gluon spin is experimentally accessible, but the
gluon OAM ${\cal L}_g^z$ is not. On the other hand, the gluon
(total) angular momentum $J_g^z$ appearing in the Ji decomposition
is accessible, either indirectly (by subtraction, using quark GPDs
from lattice QCD and/or DVCS), or directly, using by calculating
gluon GPDs on a lattice and/or deeply virtual
$J/\psi$ production. Even though $\frac{1}{2}\Delta G$ and 
$J_g^z$ belong to two
incommensurable decompositions of the nucleon spin, one may thus
be tempted to consider the difference between these two quantities,
hoping to learn something about gluon OAM. Subtracting (\ref{eq:JJM})
from (\ref{eq:JJi}), it is straightforward to convince oneself that
\be
J^z_g-\frac{1}{2}\Delta G = {\cal L}_g^z + 
\sum_q\left({\cal L}_q^z -L_q^z\right),
\ee
i.e. numerically $J^z_g-\frac{1}{2}\Delta G$ differs from 
${\cal L}_g^z$ by the same amount that $\sum_q{\cal L}_q^z$ differs 
from
$\sum_q L_q^z$. In our QED example, 
with 
\be
\Delta \gamma=\int_0^1 dx\int \frac{d^2{\bf k}_\perp}{16\pi^3}
\left[ \left|\psi_{+\frac{1}{2},+1}^\uparrow\right|^2
-\left|\psi_{+\frac{1}{2},-1}^\uparrow\right|^2 +
\left|\psi_{-\frac{1}{2},+1}^\uparrow\right|^2\right]
\ee
being the photon spin contribution, 
one thus finds (for $\lambda\rightarrow 0$, 
$\Lambda\rightarrow \infty$)
\be
J^z_\gamma-\frac{1}{2}\Delta \gamma = 
{\cal L}_\gamma^z + \frac{\alpha}{4\pi}.
\ee
As was the case in (\ref{eq:LLQED}), $\frac{\alpha}{4\pi}$ appears
to be a small correction, but one needs to keep in mind that  for
an electron
$J^z_\gamma$, $\Delta \gamma$, and ${\cal L}_\gamma^z$ are also
only of order $\alpha$.

\section{Discussion and Summary}

We have studied both the Jaffe/Manohar, as well as the Ji
decomposition of angular momentum in the scalar diquark model, as
well as for an electron in QED to order $\alpha$. As expected,
both decompositions yield the same numerical value for the fermion
OAM in the scalar diquark model, but not in QED. This calculation
demonstrates explicitly
that the presence of the vector potential in the
manifestly gauge invariant local operator for the OAM does indeed
contribute significantly to the numerical value of the OAM.
While the numerical value for difference between the fermion OAM
in these two decompositions in QED appears to be small 
($\frac{\alpha}{4\pi}$), one should keep in mind that the OAM itself
is of the same order $\alpha$. Moreover, applying the same calculation
to a massive quark in QCD yields a contribution 
from the vector potential term to the angular momentum of the quark
of about $-10\%$ (for $\alpha_S\approx 0.5$). 

%The calculations were done using explicit expressions for the
%2-particle Fock components of the light-cone wave functions
%in these systems (the orbital angular momentum in the 1-particle
%Fock component is zero). A direct evaluation of the matrix elements 
%of 
%$e q^\dagger \left[{\vec r}\times{\vec A}\right]^zq$ would receive
%nonzero contributions only from the overlap between Fock components
%that differ by one photon. To order $\alpha$, the only contributions
%would come from matrix elements of this operator between the $e$ and
%the $e\gamma$ Fock components (in such a calculation, one power of 
%the charge $e$ would come from the wave function of the $e\gamma$ Fock
%component and the other from the factor $e$ multiplying the operator).
%However, we did not perform a direct evaluation of this matrix
%element, which requires a very careful treatment involving wave 
%packets, due to the presence of the factor ${\vec r}$.
%Calculating the OAM in the Ji-decomposition
%via GPDs allowed to avoid a direct evaluation of these 
%off-diagonal matrix elements.

The sign of the contribution to the angular momentum 
arising from the vector potential is also 
significant
in light of recent lattice results for the contributions from the
$u$ and $d$ quark OAM to the nucleon spin \cite{lattice}, yielding
$L_u^z<0$ and $L_d^z>0$. The signs of the lattice results are thus
exactly opposite to what one 
would have expected on the basis of relativistic quark models, such
as the bag model, where the OAM arises from the lower Dirac component
and its expectation value is thus positively correlated to the 
expectation value of the quark spin. While the lattice results
still neglect insertions of the operator into disconnected quark
loops, this does not affect $L_u^z-L_d^z$, and the sign of that 
difference should be reliable. In Ref. \cite{tony}, evolution
has been proposed to explain this apparent discrepancy, as
a quark acquires OAM in the direction opposite to its spin
from virtual gluon emission (see Fig. {\ref{fig:cutoff}).
Our result adds to that effect in the sense that the
vector potential also adds a contribution to the OAM that is
in the opposite direction from the quark spin. Such a
shift would imply ${\cal L}_u^z>L_u^z$ and ${\cal L}_d^z<L_d^z$,
moving ${\cal L}^z_q$ closer to the quark-model-based intuitive
expectation than $L^z_q$.

%\section{Spectator Effects}
%The above studies in QED should provide some insight into the
%effect from self-generated vector potential on the manifestly
%gauge invariant OAM. However, the vector potential that enters
%the OAM operator can also arise from spectator currents. In order to
%develop some intuition about the sign/magnitude of this contribution,
%we consider in the following the bag model as a prototype.

{\bf Acknowledgements:}
M.B. would like to thank A. Bacchetta, S.J. Brodsky, and M. Diehl for 
useful 
comments.
This work was supported by the DOE under grant number 
DE-FG03-95ER40965 and (M.B.) DE-AC05-06OR23177 (under which Jefferson 
Science Associates, LLC, operates Jefferson Lab). 

\bibliography{oam.bbl}% Produces the bibliography via BibTeX.
\end{document}